# Interpretability of artificial neural network models in artificial intelligence vs. neuroscience

Kohitij Kar[*, 1, 2], Simon Kornblith[3] and Evelina Fedorenko[1]

[1]McGovern Institute for Brain Research and Department of Brain and Cognitive Sciences, Massachusetts Institute of Technology, Cambridge, MA, 01239, USA
[2]Center for Brains, Minds, and Machines, Massachusetts Institute of Technology, Cambridge, MA, 01239, USA
[3]Google Research, Brain Team
[*]Correspondence should be addressed to Kohitij Kar (kohitij@mit.edu)

## Summary

Computationally explicit hypotheses of brain function derived from machine learning (ML)-based models have recently revolutionized neuroscience [1,2]. Despite the unprecedented ability of these artificial neural networks (ANNs) to capture responses in biological neural networks (brains) (**Fig. 1A**; see [3] for a comprehensive review), and our full access to all internal model components (unlike the brain), ANNs are often referred to as "black boxes" with limited interpretability. Interpretability, however, is a multi-faceted construct that is used differently across fields. In particular, interpretability, or explainability, efforts in Artificial Intelligence (AI) focus on understanding how different model components contribute to its output (i.e., decision making). In contrast, the neuroscientific interpretability of ANNs requires explicit alignment between model components and neuroscientific constructs (e.g., different brain areas or phenomena, like recurrence [4] or top-down feedback [5]). Given the widespread calls to improve the interpretability of AI systems [6], we here highlight these different notions of interpretability and argue that the neuroscientific interpretability of ANNs can be pursued in parallel with, but independently from, the ongoing efforts in AI. Certain ML techniques (e.g., deep dream, see **Fig. 1C**) can be leveraged in both fields, to ask what stimulus optimally activates the specific model features (feature visualization by optimization), or how different features contribute to the model's output (feature attribution). However, without appropriate brain alignment, certain features (non-blue segments of the model in **Fig. 1C**) will remain uninterpretable to neuroscientists.

## A conceptual framework for operationalizing ANN interpretability in neuroscience

Like interpreters of human languages, scientists seek a high-fidelity mapping between two "languages": the "language" of scientific measurement, and the "language" of scientific hypotheses (models). The language of measurement consists of numerical descriptions of a sample from the phenomenon we seek to understand. For instance, systems neuroscientists interested in visual processing could measure and summarize neural spiking activity (e.g., time-averaged firing rates) from individual visuocortical neurons or obtain behavioral measurements (e.g., task accuracies, or reaction times) on specific visual tasks. The language of scientific hypotheses consists of conceptual abstractions that aim to explain, predict, and control the phenomenon of interest (e.g., the pattern of firing rates across neurons predicting the category of visual objects [7] in parts of the ventral visual stream). To claim that a specific model (or parts of it) is uninterpretable to a neuroscientist then means that certain components/features of the model do not map onto any empirically verifiable neuroscientific construct. To our knowledge, all current ANN models of primate vision [3] contain features that have not been explicitly mapped onto neuroscientific constructs, limiting their interpretability (see an example depiction of such a scenario in **Fig. 1C**).

## Interpretability lies in the eyes of the interpreter

We further propose that neuroscientific interpretability is itself a relative term. In particular, the extent to which a model needs to be accessible to a human experimenter (allowing for easy comparisons with empirical data), and aligned with neuroscientific constructs relies on the model's intended use. For instance, a model that is expected to predict the responses in the fMRI-based voxels in specific subregions of the human brain need not map onto the lower-level components of the brain like the neurons. It should, however, have explicit mapping onto all accessible experimental components like the ability to engage with the exact stimulus and the ability to perform the behavioral task. Therefore, a model that is interpretable for one set of experiments may not be interpretable for another. This grain-dependent interpretability, however, should not necessarily discourage models to comprise finer-grain details of the brain but will require the modeler to minimally commit to a clear mapping between model features and specific relevant experimental variables of interest.

## Leveraging the synergies – Neuroscientific interpretability as a benchmark for the goodness of ANN explainability

There has been growing interest and legislation across many leading nations of AI research to promote and achieve explainable AI [6,8]. However, no ground truth exists for what constitutes a good explanation. Indeed, one of the significant challenges for the current AI "explainability" results is that different methods to interpret the functional role of model features lead to different results and inferences [9], and it is unclear how to evaluate explanation quality. We propose that one way to validate the "goodness" of explanations is to measure their match with the explanations derived from primate behavior [10,11] and neural measurements [12,13]. We are not necessarily proposing that the brain is an optimal system that AI should mimic. But given that the errors made by top-performing machine learning models increasingly resemble those made by humans [14], it is reasonable to expect some degree of mechanistic alignment. Thus, in the absence of any ground truth in AI explainability, we argue that similarity with the primate brain (a system that is robust, flexible, and capable of powerful generalization) might provide valuable guidance. Quantitatively assessing alignment between ML feature attribution measures and factors that are critical for primate decision-making and brain activation could therefore serve as a putative benchmark for AI explainability measures.

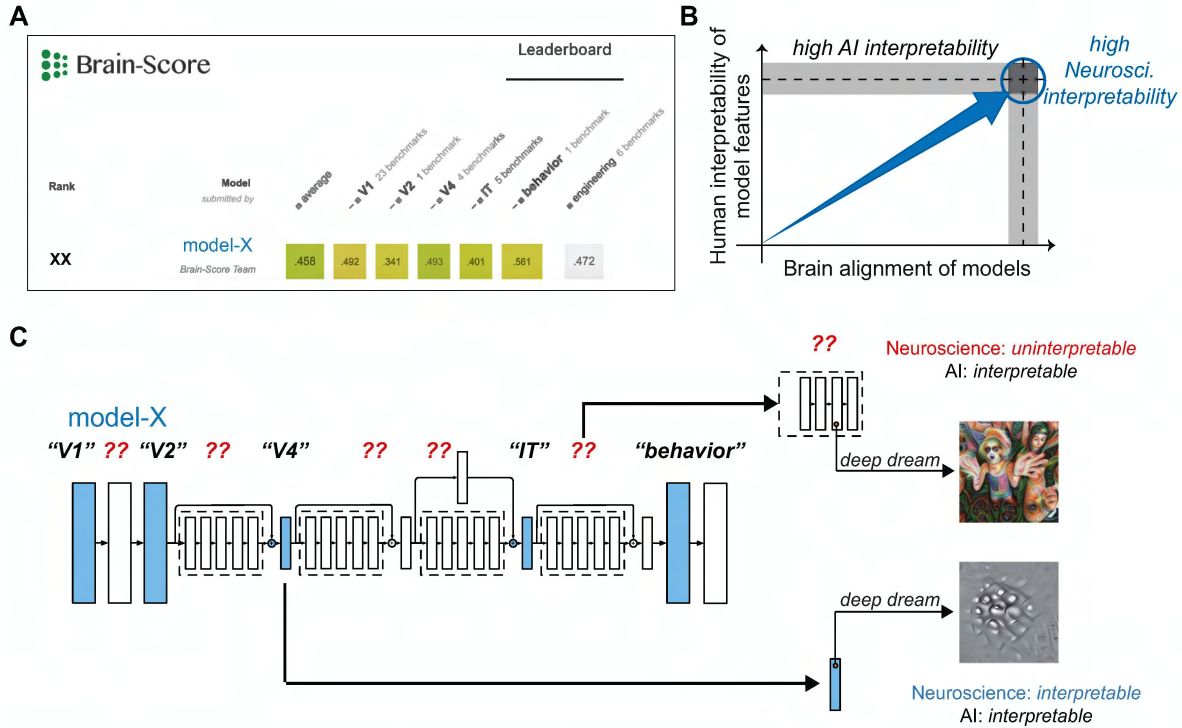

**Figure 1** Interpretability of models for AI and Neuroscience. **A.** A schematic depiction of the Brain-Score platform[3] website. Columns contain scores for different brain benchmarks (e.g. V1, V2, V4, IT predictivity), and each row corresponds to a specific model (example shown here as a dummy, model-X). **B.** How brain alignment of models and human interpretability of model features (and their attributes) relate to high AI and neuroscientific interpretability of models. For instance, higher levels of human interpretability of feature attribution might make ANNs highly interpretable for AI (refer to the top left part of the plot), but poor brain alignment will lead to low interpretability for the neuroscientists. On the other hand, higher brain alignment and a high level of human interpretability will lead to models that are more interpretable for both AI and the neurosciences (refer to the top right part of the plot). **C.** Schematic of model-X (this could be replaced by almost any high-ranking model from Brain-Score[3]). We show an example of how only some parts (in blue) of the model can be interpretable to a neuroscientist (for example see[12]) since they can directly map to brain areas. As we show here, ML tools (like deep dream) could exist to interpret the entire model for AI (e.g. both blue and white boxes).

## Conclusion

To conclude, we encourage researchers to conceptually separate the objectives of AI and neuroscience while interpreting the parameters and operations of current computational models. However, we also suggest that—although current AI models operate differently from primate brains in various ways—all else being equal, interpretability methods in AI that provide more primate-brain-aligned model interpretations are likely to be more promising.

## Acknowledgments

The authors would like to thank Cory Shain for helpful comments and discussions.

## Funding

EF was supported by NIH awards R01-DC016607, R01-DC016950, and U01-NS121471, and by research funds from the McGovern Institute for Brain Research, the Brain and Cognitive Sciences Department, and the Simons Center for the Social Brain.